\documentclass[12pt]{iopart}
\usepackage{epsfig}
\newcommand{\be}{\begin{equation}}
\newcommand{\ee}{\end{equation}}
\newcommand{\ba}{\begin{eqnarray}}
\newcommand{\ea}{\end{eqnarray}}
\newcommand{\bdm}{\begin{displaymath}}
\newcommand{\edm}{\end{displaymath}}
\begin{document}
\title{Exact Results for a Spin-${\bf 1}$ lattice}

\author{Sungkit Yip\dag
\footnote[3]{
%To whom correspondence should be addressed ( 
yip@phys.sinica.edu.tw}
}

\address{\dag
Institute of Physics, Academia Sinica, Nankang, Taipei 115,
Taiwan}
\date{\today}

\begin{abstract}
We consider a lattice of spin-$1$ particles with
a general pairwise interaction 
$ \left[ \ {\rm cos} \gamma \ ( {\bf S}_{l} \cdot {\bf S}_{l+1} ) \
   + \ {\rm sin} \gamma \ ( {\bf S}_{l} \cdot {\bf S}_{l+1} )^2 \ 
   \right]$.
We show that, for a large class of lattices with even number 
of sites, the ground state for the region
$ - { 3 \pi \over 4} < \gamma < - {\pi \over 2}$
belongs to total spin $S_{\rm tot} = 0$, whereas
the state of minimum excited energy but with finite $S_{\rm tot}$
belongs to $S_{\rm tot} = 2$.
These results are constrasted with the generalized
Marshall theorems, applicable to 
a bipartite lattice and $ - {\pi \over 2} < \gamma \le 0$.
\end{abstract}

%Uncomment for PACS numbers title message
\pacs{75.10.-b, 75.10.Jm, 75.10.Pq}

% Uncomment for Submitted to journal title message
\submitto{\JPC}

% Comment out if separate title page not required
\maketitle

\section{Introduction}

In this paper, we consider a lattice
consisting of ${\cal N}$ spin $ S = 1$, interacting with each other
through a pairwise interaction.
An example is the Hamiltonian for a one-dimensional spin-$1$ chain
with nearest neighbor interaction,

\ba
{\cal H}  &=& \sum_l \ \left[ 
   \ J \ ( {\bf S}_{l} \cdot {\bf S}_{l+1} )
  \  + \ K \ ( {\bf S}_{l} \cdot {\bf S}_{l+1} )^2 \ \right]
\label{H} \\
&\equiv& \sqrt{J^2 + K^2} \  
\sum_l \ \left[ \ {\rm cos} \gamma \ ( {\bf S}_{l} \cdot {\bf S}_{l+1} ) \
   + \ {\rm sin} \gamma \ ( {\bf S}_{l} \cdot {\bf S}_{l+1} )^2 \ 
   \right]
\label{gamma}
\ea
where the sum is over the site labels $l$ and
the second relation defines $\gamma$. 
This Hamiltonian has been of intense interest
in the theoretical studies 
 on quantum magnetism \cite{review,Parkinson,Fath93,Fath95}
Here we shall demonstrate some exact properties
of the ground states for this type of Hamiltonians.
For definiteness, we shall confine ourselves
to the Hamiltonian (\ref{H}) (or its slight generalization (\ref{H1a})
below) with
${\cal N}$ being even and finite unless otherwise stated.
We shall be particularly interested in the region
 $ K < J < 0$, {\it i.e.},  
$ - { 3 \pi \over 4} < \gamma < - {\pi \over 2}$.

In a recent paper \cite{Yip03}, we have pointed out
that this region of $\gamma$ is relevant to
spin-$1$ Bosons trapped in an (optical) lattice in
the regime of one particle, one orbital state per site,
for suitable interactions between the Bosons.
To see this, we generalize the standard
derivation of the Heisenberg (exchange) Hamiltonian from the Hubbard model
\cite{Auerbach94} to spin $1$.   
Let the hopping between sites be $t$ and
the (repulsive) interaction for two particles on the same site be
$U_0 (>0) $ if their total spin is $0$, and 
$U_2 (>0) $ if their total spin is $2$.
(Note that they cannot have total spin $1$ since we have identical
Bosons and there is only one orbital per site.)
Consider now two neighboring sites (say $1$ and $2$) with one Boson each.
At $t =0$ the states corresponding to different spin configurations
are degenerate.  For finite but small $t$ we can perform 
perturbation to second order.  The energies of the system
can be classified according to the total spin $S_{\rm tot}$ for the two
sites, and can be easily seen to be 
 $ - 4 t^2 / U_{0}$, $0$, and $- 4 t^2 / U_{2}$ for total spin 
$S= 0, 1, 2$ respectively.  Comparing these values with 
those of the general form of the spin Hamiltonian  
$\epsilon_0 + J ( {\bf S}_1 \cdot {\bf S}_2 )
    \  +  \ K ( {\bf S}_1 \cdot {\bf S}_2 )^2$
for these two sites, and noting that 
${\bf S}_1 \cdot {\bf S}_2 = -2, -1, 1$ 
for $S_{\rm tot} = 0, 1, 2$ respectively,
we find
$J = - { 2 t^2 \over U_2}$, $K = - { 2 \over 3} { t^2 \over U_2}
        - { 4 \over 3} { t^2 \over U_0 } $
and $\epsilon_0 = J - K $.  Thus $ K < J < 0$ if $0 < U_0 < U_2$.
($\epsilon_0$ represents an energy shift independent of 
the spin configurations and can thus be dropped from our discussions.)
The case $ 0 < U_0 < U_2$ is applicable to $^{23}$Na atoms.
\cite{Ho00}

  Our major results are collected as {\it Theorems} and {\it Corollaries}
in Sec \ref{sec:na}.  In particular, we shall show that the
ground state for the Hamiltonian (\ref{H}) has zero total spin
$S_{\rm tot} = 0$.  We also show that the excitation of lowest energy
but with finite spin has $S_{\rm tot} = 2$.
Generalizations of our results to lattices other than one-dimensional
chains are also possible.  Some of them are mentioned
in subsection \ref{subsec:na2}.

For comparison, we shall, in section \ref{sec:rb}, discuss the
case where 
 $ J <  K  <  0$  [$  - {\pi} < \gamma < - { 3 \pi \over 4} $].
This case applies to spin-$1$ Bosons in a lattice
if the interaction among Bosons is "ferromagnetic",
($ U_2 < U_0$)
as would be the case for $^{87}$Rb atoms trapped in an optical
lattice \cite{Yip03}.  It is worth remarking here that
spin-polarized $^{87}$Rb in an optical lattice in the
(Mott) regime of one particle per site has already been
obtained in a recent experiment \cite{Greiner02}. 

Our work generalizes the corresponding results for the 
Heisenberg ( $ J \ne 0 $, $ K = 0$ ) Hamiltonian.
For a bipartite lattice and $J > 0$ (the antiferromagnetic Heisenberg
model), the results are usually referred to as the
Marshall theorems \cite{Auerbach94}.  The proofs here follow
similar ideas, but the details are very different 
(in particular, our results here do not rely on a bipartite lattice).
Section \ref{sec:com} below contains a comparison between the two cases.
Main results are summarized in section \ref{sec:concl}.

\section{The region $K < J < 0$:}\label{sec:na}

To ease the presentation we shall first consider a 
linear spin chain (subsection \ref{subsec:na1}).
Then we shall discuss more general lattice types
(subsection \ref{subsec:na2})

\subsection{Linear spin chain}\label{subsec:na1}

\begin{figure}
\begin{center}
{\epsfbox{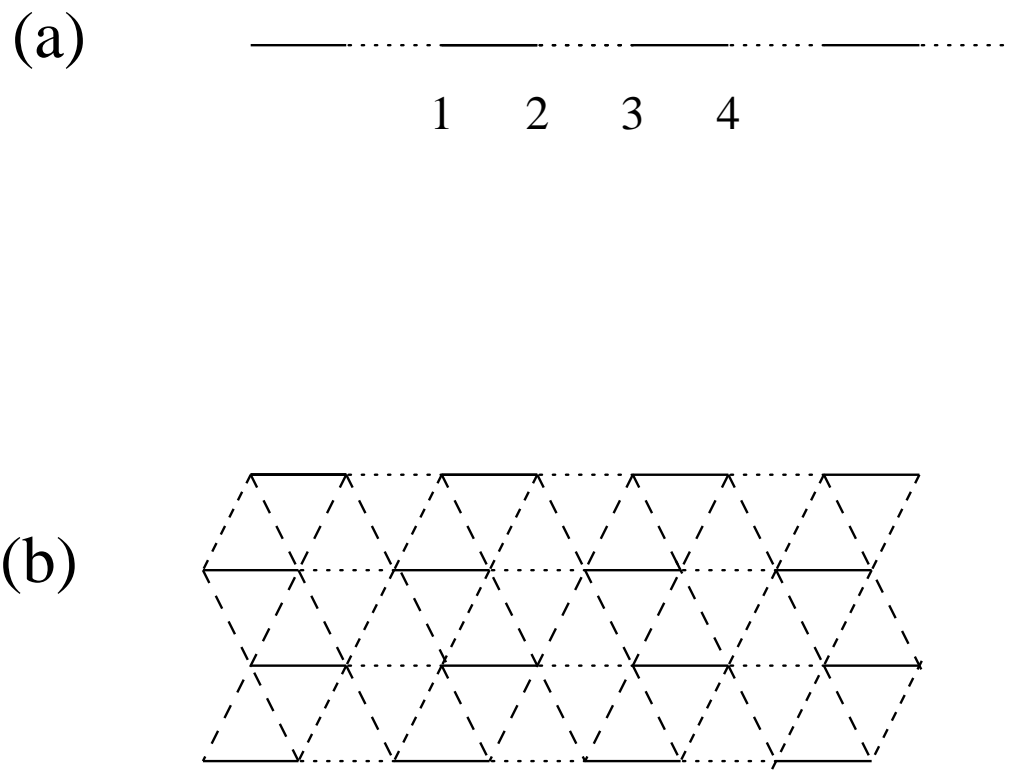}}
\end{center}
\caption{\label{fig:al} (a) Schematic representation of the spin $1$
chain with Hamiltonian (\ref{H1a}).
Thick (dotted) lines represent stronger (weaker) bonds,
with strengths proportional to $1 + \delta$ ( $ 1 - \delta$).
(b) Schematic representation of a hexagonal spin $1$ lattice.}
\end{figure}

We shall consider, for definiteness,
 a generalized form of eq (\ref{H}):

\be
\hat H = \sum_l \  \left( 1 - ( -1 ) ^l \delta \right) \left[ 
   \ J \ ( {\bf S}_{l} \cdot {\bf S}_{l+1} )
  \  + \ K \ ( {\bf S}_{l} \cdot {\bf S}_{l+1} )^2 \ \right]
\label{H1a}
\ee
where $ 0 \le \delta \le 1 $, {\it i.e.}, with
bond alternation ( in strength but not in sign of the interaction).
This is shown schematically in Fig \ref{fig:al}.
For convenience we shall also define $\epsilon = 1 - \delta \ge 0$.
$\epsilon$ is thus proportional to the strength of the weaker
bonds.
For $\delta = 0$ ($\epsilon = 1$),
Hamiltonian (\ref{H1a}) reduces to that of (\ref{H}).
For $\delta = 1$ ($\epsilon = 0$), the Hamiltonian can
be solved trivially since we have a collection of spin
pairs with no interaction among different pairs,
a fact that we shall take advantage of.

We shall use the `Ising configurations'
as our basis set.  Each of these configurations
is defined by specifying the spin projection
along the $\hat z$ direction for every site.
Denoting the state at a site by $ |+>, |0>, | - > $
according to $S_z = 1, 0, -1$, a typical Ising
configuration, which will be represented by 
$| \alpha > $,  is, e.g., $ | +  - 0 + - - 0 0 ... > $ etc.
We shall express any many body state $|\Psi>$
as a linear combination of these basis states,
{ \it i.e} 
\be
 | \Psi > = \sum_{\alpha} c_{\alpha} | \alpha > 
\label{exp1}
\ee
where the sum is over all $|\alpha> $.
We shall also introduce the rotated basis,
defined by, for {\it all} sites,
\be
  \begin{array}{cccc}
 | + >' &=&    & | + > \\
 | - >' &=&    & | - > \\
 | 0 >' &=&  i & | 0 > 
\label{rot}
  \end{array}
\ee
and also use the expansion
\be
 | \Psi > = \sum_{\alpha'} c_{\alpha'} | \alpha >' 
\label{exp2}
\ee
The factor $i$ in eq (\ref{rot}) should not alarm us.
Since the $z$ component of the total spin 
$M \equiv S_{{\rm tot},z}$ is conserved under ${\cal H}$, for each 
given $M$, the number of sites $l$ with $S_{z,l} = 0 $ 
in the expansions (\ref{exp1}) or (\ref{exp2})
must be either even
or odd for all $\alpha$'s [ for given $M$, 
the number of these sites can be only changed by replacing  $ + - $ pairs ( not
necessarily nearest neighbours) by $ 0 0 $, or vice versa ]. 

\noindent {\it Theorem 1}:  For the lowest energy state of ${\cal H}$ 
of a given $M$,  
the coefficients $c_{\alpha'}$ are
non-negative for all $|\alpha>'$  (apart from an overall common phase
factor, a caveat that we shall not repeat).

\noindent {\it Proof}:  The basic idea of the proof is similar to that
of the Marshall theorem \cite{Auerbach94}.  One regards
the Schr\"odinger equation 
$ {\cal H} | \Psi > = \ E \ | \Psi > $ as the corresponding one
for a tight-binding (hopping) Hamiltonian with the 
lattice points labelled by $ |\alpha >$ ( $|\alpha >'$ ).
Consider the general term in the Hamiltonian between sites 
$l$ and $l+1$:
\bdm
 H_{l,l+1}  = \left( 1 \pm \delta \right) \times \left[ 
   \ J \ ( {\bf S}_{l} \cdot {\bf S}_{l+1} )
  \  + \ K \ ( {\bf S}_{l} \cdot {\bf S}_{l+1} )^2 \ \right]
\edm
This term affects only the spins $l$ and $l+1$.  When
$H_{l,l+1}$ operates on an Ising configuration $|\alpha>$,
spin configurations for $l' \ne l, l+1$ are unchanged.
For simplicity, we shall suppress the configurations for all these
$l'$.  For the spin configurations of $l$ and $l+1$, $H_{l,l+1}$ 
only makes `hopping' among $ | + - >$, $| 0 0 >$, $ | - + > $;
between $ | + 0 > $ and $ | 0 + > $,
and between   $ | - 0 > $ and $ | 0 - > $.
In the first mentioned subspace 
 ( $ | + - >$, $| 0 0 >$, $ | - + > $),
 the matrix elements of
$H_{l,l+1}$ are given by,
\be
   ( 1 \pm \delta ) \times \left(
  \begin{array}{ccc}
   - J + 2 K &  J - K & K      \\
   J - K &     2 K    &  J - K \\
     K   &   J - K    &  - J + 2 K 
   \end{array}
   \right)
\label{3d}
\ee
In the second mentioned subspace 
( $ | + 0 > $ and $ | 0 + > $ ), they are
\be
  ( 1 \pm \delta ) \times \left(
  \begin{array}{cc}
    K &    J      \\
   J &     K 
   \end{array}
   \right)
\label{2d}
\ee
with identical matrix elements for the third 
subspace (  $ | - 0 > $ and $ | 0 - > $ ).
 $H_{l,l+1}$ is diagonal in $ | + + >$ and $ |- - >$.

In the rotated basis, the matrix (\ref{3d}) transforms to
\be
   ( 1  \pm \delta ) \times
  \left(
  \begin{array}{ccc}
   - J + 2 K &  - (J - K) & K      \\
   -(J - K) &     2 K    &  - (J - K) \\
     K   &   -(J - K)    &  - J + 2 K 
   \end{array}
   \right)
\label{3da}
\ee
while the matrix (\ref{2d}) is unchanged. In our $\gamma$
region of interest, $J - K > 0$, $K < 0 $, $J < 0$,
hence all hopping (off-diagonal) matrix elements 
are $\le 0$.  It follows that the lowest energy
state must have $c'_{\alpha} \ge 0$ for all $|\alpha>'$.

\noindent{\it Corollary $1$}:  If $\delta \ne 1$, the lowest energy
state for given $M$ has $c_{\alpha'} > 0$ for all $\alpha'$'s.
  It also follows that this state is unique for given $M$.

\noindent{\it Proof}: If $\delta \ne 1$ ($\epsilon > 0$), all hopping matrix
elements are negative (non-zero).  As in the case of
the Heisenberg Hamiltonian \cite{Auerbach94}, all Ising configurations
are connected by (though multiple) hopping.  By the same
reasoning as in \cite{Auerbach94}, the lowest energy
state must have all coefficients 
$c_{\alpha}'$ non-zero; and moreover there can be only
one such state.

\noindent{\it Remarks}:  For $\delta = 1$ ($\epsilon = 0$) the system
becomes a collection of spin pairs, with no interaction
among pairs.  For one pair, the energies $E_{S}$ only
depends on the total spin $S$ and are given by
\ba
E_0 &= 2 \left[ - 2 J + 4 K \right] \nonumber \\
E_1 &= 2 \left[ - \ J  + \ K \right]  \\
E_2 &= 2 \left[ \ J + \ K \right] \nonumber
\label{epair}
\ea
For $K < J < 0$, $E_0 < E_2 < E_1$.
The minimum energy state is
given by, for $M = 0$,
${ 1 \over {\sqrt 3} } \left[ | + - > + | - + > - | 0 0 > \right] $
$= $ 
${ 1 \over {\sqrt 3} } \left[ | + - >' + | - + >'  + | 0 0 >' \right] $;
for $M =1 $, 
${ 1 \over {\sqrt 2} } \left[ | + 0 > + | 0 + >  \right] $
$= $ 
$ - { i \over {\sqrt 2} } \left[ | + 0 >' + | 0 + >'  \right] $.
and for $M = 2$,
$ | + + > = | + + >'$.  The first state
has $S=0$ and the latter two both belong to $S=2$.
These states obviously obeys the signs stated in {\it Theorem} $1$.

For ${\cal N} = 2 N$ spins, the ground state has $S_{\rm tot} = M = 0$
and is a collection of singlet pairs.  For $M=1$ ($2$),
the lowest energy states have one pair of spins in the
$S = 2$, $S_z = 1$ ($2$) state with the rest in the singlet states.
These states are not unique due to the freedom of choice
of which pair being the $S=2$ pair.

\noindent {\it Theorem} $2$:  For a given $M$, the lowest energy state
for $0 \le \delta \le 1$ has the same $S_{\rm tot}$
independent of $\delta$.

\noindent {\it Proof}: For any given $\delta \ne 1$ we have
already seen that the minimum energy state has
$c_{\alpha'}(\delta) > 0$ for {\it all}
Ising configurations $\alpha'$.  For $\delta = 1$,
we have $c_{\alpha'}(1) \ge 0$.  Their overlap, given
by $\sum_{\alpha'} c_{\alpha'}(1) \times c_{\alpha'}(\delta)$,
is non-vanishing.  Hence they must have the same $S_{\rm tot}$.

 From the {\it Remarks} following
{\it Corollary} $1$, it follows that, for our region of $\gamma$,

\noindent (i) The lowest energy state in the $M=0$ sector has $S_{\rm tot} = 0$,

\noindent (ii) the lowest energy state in the $M=1$ sector and 
the lowest energy state in the $M=2$ sector
both have $S_{\rm tot} = 2$.

Since the $M$ sector contain states with $S_{\rm tot} \ge M$, we have

\noindent {\it Corollary} $2$:

\noindent (a) The ground state for our region 
($ K < J < 0 $) has $S_{\rm tot} = 0$

\noindent (b) The spin excitation  ($S \ne 0$) with the smallest
excited energy belongs to $S_{\rm tot} = 2$.

\noindent (c) Any state with $S_{\rm tot} = 1$ has energy higher than 
the mentioned state in (b) if $\delta \ne 1$ ($\epsilon > 0$)

  The lines of reasoning above can also applied to deduce some
properties concerning states of large $S_{\rm tot}$.
The sector $M = 2 N - 1$ has one state belonging to $S_{\rm tot} = 2N$
and one belonging to $S_{\rm tot} = 2 N -1$.
The former state can trivially be written down and 
has $c_{\alpha'} \ge 0$.  Hence we conclude that
$E_{2N} \le E_{2N-1}$, and this later expression becomes
a strict inequality if $\epsilon \ne 0$.

For odd ${\cal N}$, similar considerations above show that
the ground state has $S_{\rm tot} = 1$.
 
The conclusions in this Corollary agree with known
numerical results ({\it e.g.} \cite{Fath93}).

%This concludes the discussions for $K < J < 0$.

\subsection{Other lattice types}\label{subsec:na2}

As may already have been obvious, our {\it Theorems}
and {\it Proofs} in the last subsection 
can be generalized to other lattice types.
Moreover, these lattices need not be bipartite.
A particular interesting case is as shown in Fig \ref{fig:al}b.
Here we are again considering a spin-$1$ lattice with pair-wise
interactions  
$ J ( {\bf S}_1 \cdot {\bf S}_2 )
    \  +  \ K ( {\bf S}_1 \cdot {\bf S}_2 )^2$
with the strengths of the bonds proportional to, for full lines, $1$,
dashed lines, $ \epsilon_1$, and dotted lines, $\epsilon_2$.
($\epsilon_1,\epsilon_2 \ge 0$ but not necessarily equal.)
The case where all bond strengths are equal is contained 
in $\epsilon_1 = \epsilon_2 = 1$. 
The arguments leading to {\em Theorem 1} and {\it Corollary 1} went through
unchanged.  
 If $\epsilon_1 = \epsilon_2 = 0$, the system again reduces
 to the collection of pairs, and the {\it Remarks}
subsection following {\it Corollary 1} applies.
By comparing the system of interest
(with $\epsilon_1$, $\epsilon_2$ both finite)
with that with $\epsilon_1 = \epsilon_2 = 0$, we can again
prove {\it Theorem} $2$ and {\it Corollary} $2$ as before.
It is worth mentioning that a triangular optical lattice
can also be formed by three suitable laser beams
propagating at angles $ \pi /3 $ with respect to each other,
thus the discussions here is applicable also to 
a physical system.

\section{The region $J < K < 0$ }\label{sec:rb}

Similar arguments above can easily be generalized to the region
$J < K < 0$ ($ - \pi < \gamma < - { 3 \pi \over 4}$)
for Hamiltonian (\ref{H1a}).  In this case with $0 \le \delta < 1$,
all hopping matrix elements are negative in the 
original (unprimed) basis, and thus the ground state for
each $M$ sector has $c_{\alpha} > 0$.
  This result is in agreement with the common wisdom expressed
in the literature ({\it e.g.}, Ref \cite{Fath95}) that
the ground state here has $S_{\rm tot} = 2 N$.  These
states, obtainable from $| ++++...>$ by suitable number of
lowering operators, indeed has $c_{\alpha} > 0$ for all $|\alpha>$'s.

\section{Comparison with Marshall theorems}\label{sec:com}

For the antiferromagnetic Heisenberg ($J > 0$, $K = 0$) model
{\it on a bipartite lattice}, the Marshall theorems hold.
The basis employed in these theorems is defined by,
for the A sublattice,
\ba
| + > _{\cal M} &=& | + > \nonumber \\
| 0 > _{\cal M} &=& | 0 > \\
| - > _{\cal M} &=& | - > \nonumber 
\ea
and for the B sublattice,
\ba
| + > _{\cal M} &=& - | + > \nonumber \\
| 0 > _{\cal M} &=& \ | 0 > \\
| - > _{\cal M} &=& - | - > \nonumber
\ea

For our more general Hamiltonian (\ref{H1a}), the matrix elements
of $H_{l,l+1}$ in this basis is given also by (\ref{3da})
for the  $ | + - >$, $| 0 0 >$, $ | - + > $ subspace while
that for $ | + 0 > $ and $ | 0 + > $ (or $ | - 0> $ and $ | 0 - >$)
becomes 
\be
  ( 1 \pm \delta ) \times \left(
  \begin{array}{cc}
    K &   - J      \\
  - J &     K 
   \end{array}
   \right)
\label{2dM}
\ee

Thus, for $J > 0$, $K < 0$  ($ J-K > 0$ trivially) all "hopping"
elements are negative.  Thus the Marshall theorems can be
generalized to the region $ - {\pi \over 2} < \gamma < 0$
with the results (arguing as in \cite{Auerbach94} or as in
Sec \ref{sec:na}; note that in this region, for a system with
two spins, one has $E_0 < E_1 < E_2$)

\noindent (a${\cal M}$) The ground state of (\ref{H1a}) has $S_{\rm tot} = 0$
and is unique ( if $ \delta \ne 1$)

\noindent (b${\cal M}$) The spin excitation with lowest energy has 
  $S_{\rm tot} = 1$.

\noindent (c${\cal M}$) Any state with $S_{\rm tot} = 2$ has energy
higher than the state mentioned in (b${\cal M}$) if $\epsilon > 0$. 

For $\gamma = - { \pi \over 2}$, some hopping matrix elements 
vanish ($J = 0$).  There are in general degeneracies for the 
minimum energy states for a given $M$.  This fact can also
be seen from the work of Parkinson \cite{Parkinson}.

\section{Concluding remarks}\label{sec:concl}

We have proven some exact properties of lowest energy states
for Hamiltonian of the type (\ref{H1a}) with $K < J < 0$.
We demonstrated that, for a large class of lattices 
(not necessarily bipartite) with even number 
of sites, the ground state
belongs to total spin $S_{\rm tot} = 0$, whereas
the state of minimum excited energy but with finite $S_{\rm tot}$
belongs to $S_{\rm tot} = 2$.
The results derived here are relevant to spin-$1$ Bosons
trapped in an optical lattice in the regime of 
one particle per site for suitable interaction between the Bosons.

\section*{References}

%\begin{harvard}

%\end{harvard}

%%%%%%%%%%%%%%%%%%%%%%%%%%%%%%%%%%%%%%%%%%%%%%%%%%%%%
%

\clearpage

\end{document}